\documentclass[aps,pre,twocolumn,reprint,showpacs,superscriptaddress,floatfix]{revtex4-1}
\usepackage{amstext,amssymb}
\usepackage{amsmath}
\usepackage{indentfirst}
\usepackage{makeidx}
\usepackage{slashed}
\usepackage{graphicx}
\usepackage{xcolor}

\newcommand{\sm}{\scriptscriptstyle}               
\newcommand{\be}{\beta}                            
\newcommand{\al}{\alpha}                           
\newcommand{\ch}{\mathrm{C}}                       
\newcommand{\dm}{\rho}                             
\newcommand{\dk}{\gamma}                           
\newcommand{\dkz}{\gamma_{\sm{0}}}                           
\newcommand{\ev}{\mathrm{U}}                       
\newcommand{\Hz}{H_{\sm{\mathrm{o}}}}              
\newcommand{\h}{\hbar}                             
\newcommand{\id}{\mathbb{I}}                       
\newcommand{\p}{\prime}                            
\newcommand{\drm}{\mathrm{d}}                      
\newcommand{\e}{\mathop{\mathrm{e}}\nolimits}      
\newcommand{\iu}{i\,}                              
\newcommand{\sd}{\mathrm{J}}                       
\newcommand{\ep}{\lambda}                          
\newcommand{\tf}{\Delta}                           
\newcommand{\tp}{\otimes}                          
\newcommand{\w}{\omega}                            
\newcommand{\Wre}{\tilde{W}^{\al\prime}}              
\newcommand{\Wim}{\tilde{W}^{\al\prime\prime}}        
\begin{document}
\title{Reduced density matrix for nonequilibrium steady states: A modified Redfield solution approach}

\author{Juzar~Thingna}
\email[]{Electronic address: juzar.thingna@physik.uni-augsburg.de}
\affiliation{Department of Physics and Center for Computational Science and Engineering, National University of Singapore, Singapore 117542, Republic of Singapore}
\affiliation{Institut f\"ur Physik Universit\"at Augsburg, Universit\"atsstrasse 1, D-86135 Augsburg, Germany}
\author{Jian-Sheng~Wang}
\affiliation{Department of Physics and Center for Computational Science and Engineering, National University of Singapore, Singapore 117542, Republic of Singapore}
\author{Peter H\"anggi}
\affiliation{Department of Physics and Center for Computational Science and Engineering, National University of Singapore, Singapore 117542, Republic of Singapore}
\affiliation{Institut f\"ur Physik Universit\"at Augsburg, Universit\"atsstrasse 1, D-86135 Augsburg, Germany}

\date{4 November 2013}
\begin{abstract}
We describe a method to obtain the reduced density matrix (RDM) correct up to second order in system-bath coupling in \emph{nonequilibrium} steady state situations. The RDM is obtained via a scheme based on analytic continuation, using the time-local Redfield-like quantum master equation, which was earlier used by the same authors [J. Chem. Phys. \textbf{136}, 194110 (2012)] to obtain the correct thermal equilibrium description. This nonequilibrium modified Redfield solution is then corroborated with the exact RDM obtained via the nonequilibrium Green's function technique for the quantum harmonic oscillator. Lastly, the scheme is compared to different quantum master equations (QMEs), namely the time-local Redfield-like and the Lindblad-like QMEs, in order to illustrate the differences between each of these approaches.
\end{abstract}

\maketitle
\section{Introduction}
\label{sec:1}
Statistical mechanics \cite{Landau1980, Callen1985, Huang1987, Kubo1991} has been one of the cornerstones of equilibrium physics, which has enabled us to calculate any arbitrary macroscopic property of a system in contact (implicitly assumed to be \emph{weak}) with a giant environment (typically referred to as bath). Even though the framework of statistical mechanics has been highly successful, its nonequilibrium counterpart still eludes the scientific community. Even in the particular case of steady states, the nonequilibrium formulation is predominantly mathematical and formal \cite{Oono1998, Kita2002, Sasa2006} without a \emph{general} prescription to obtain the exact reduced density matrix.  

Unlike Gibbs, who searched for a statistical ensemble compatible with equilibrium thermodynamics, we focus on obtaining a nonequilibrium steady-state reduced density matrix compatible with the laws of quantum dynamics. This objective is known under a wider label of open quantum systems \cite{Carmichael1994, Blum1996, Breuer2002, Weiss2008}. The goal is typically addressed by use of a wide variety of formally exact \cite{Nakajima1958, Zwanzig1960, Fulinski1968, Hanggi1977, Grabert1977, Shibata1977} or perturbative master equations \cite{
Pauli1928, Redfield1957, Davies1974, Lindblad1976}. For the steady-state nonequilibrium scenario all of the perturbative approaches are appropriate only in the regime of vanishing system-bath coupling or in the van Hove limit \cite{Mori2008, Fleming2011, Thingna2012}; see Ref.~\cite{Spohn1978} for an elucidate exposition on this limit. Hence, it is crucial to obtain the reduced density matrix (RDM) containing higher orders of the system-bath coupling in order to extract essential information about the physical quantities related to transport.

Our main goal in this paper is to obtain the nonequilibrium \emph{steady-state reduced density matrix} for a general system connected to multiple heat baths up to $2$-nd order in the system-bath coupling. We achieve this using an analytic continuity technique, which we have introduced previously \cite{Thingna2012} for the case of thermal equilibrium. For this non-trivial steady-state nonequilibrium case exact results presently exist mainly for transporting currents across classical \cite{Rieder1967} and quantum harmonic chains \cite{Zurcher1990, Segal2003, Dhar2008, Wang2008}. It is only recently that Dhar, Saito, and H\"{a}nggi \cite{Dhar2012} have analytically obtained the exact steady-state RDM of an open system composed of harmonic oscillators. Due to the lack of exact general solutions for anharmonic systems in nonequilibrium steady-state, in this work we corroborate our result with the analytic solution of Dhar \emph{et~al.} and additionally compare to other commonly used quantum master equations (QMEs). The deficiencies in those other commonly used QMEs definitely call for an accurate approach such as ours in order to investigate nonequilibrium phenomena for general anharmonic systems. 

The paper is organized as follows: In Sec.~\ref{sec:2} we describe our model and the time-local Redfield-like quantum master equation in presence of multiple baths obtained under the weak system-bath coupling approximation. The central scheme of this paper is briefly described in Sec.~\ref{sec:3}, where we obtain the nonequilibrium modified Redfield solution. Sec.~\ref{sec:4} illustrates the numerical implementation of our scheme for the steady-state density matrix of a quantum harmonic oscillator and provides comparison with the exact nonequilibrium Green's function method, the Redfield-, and the Lindblad-like QMEs, illustrating some of the pitfalls in these commonly used techniques. In Sec.~\ref{sec:5} we summarize and propose promising further extensions.
\section{Time-local quantum master equation in presence of multiple baths}
\label{sec:2}
Our basic model is similar to that used by many researchers in the field of transport and goes under a wider label of Magalinski\u{\i}-Zwanzig-Caldeira-Leggett model \cite{Magalinskii1959, Zwanzig1973, Caldeira1981, Caldeira1983} in the field of quantum dissipation. The total Hamiltonian comprises of multiple baths and a system, reading:
\begin{align}
\label{eq:no2.1}
H_{\sm{\mathrm{tot}}}&=H_{\sm{\mathrm{S}}}+\sum_{\al}\left(H_{\al}+H_{{\sm{\mathrm{R}}}\al}+H_{{\sm{\mathrm{S}}}\al}\right),
\end{align}
where $H_{\sm{\mathrm{S}}}$ denotes the generally anharmonic system Hamiltonian. Here,
\begin{align}
\label{eq:no2.2}
H_{\al}&=\sum_{k=1}^{\infty}\left(\frac{p_{k,\al}^{2}}{2m_{k,\al}}+\frac{m_{k,\al}\,\w_{k,\al}^{2}}{2}x_{k,\al}^{2}\right),
\end{align}
describes the $\al$-th bath as an infinite collection of harmonic oscillators, each having a mass $m_{k,\al}$ and a frequency $\w_{k,\al}$. This above description of the baths being of the harmonic form is one particular choice, which helps us concretize the description below, but by no means is a restriction to our approach and one could in general choose fermionic or spin baths \cite{Shao1998, Prokofev2000, Cucchietti2005, Saito2007}. The potential renormalization term is given by 
\begin{align}
\label{eq:no2.3}
H_{{\sm{\mathrm{R}}}\al}&=\left(Y^{\al}\right)^{2}\left(\frac{1}{2}\sum_{k=1}^{\infty}\frac{c_{k,\al}^{2}}{m_{k,\al}\,\w_{k,\al}^{2}}\right),
\end{align}
where $Y^{\al}$ denotes any function of the system variables connected to the $\al$-th bath and $c_{k,\al}$ denotes the coupling constant of the $k$-th oscillator with the system operator $Y^{\al}$. The renormalization term above occurs naturally when one needs to ensure the translational invariance of the total Hamiltonian. The part
\begin{align}
\label{eq:no2.4}
H_{{\sm{\mathrm{S}}\al}}&=Y^{\al}\tp B^{\al},
\end{align}
is the system-bath coupling Hamiltonian, wherein $B^{\al}$ denotes the collective bath operator. For notational ease throughout this work we will set $\h=1$ and $k_{\sm{\mathrm{B}}}=1$. In general the above Hamiltonian permits transport in the presence of temperature difference among the baths.

Given the above Hamiltonian we first derive an equation, similar to the time-local Redfield quantum master equation \cite{Redfield1957}, which helps us deduce the reduced density matrix (RDM) of the system for such a nonequilibrium scenario. We start by the basic quantum mechanical definition of the \emph{total} density matrix at any time $t$ given by,
\begin{align}
\label{eq:no2.5}
\dm_{\sm{tot}}(t)&=\ev(t)\dm_{\sm{tot}}(0)\ev(t)^{\dagger},
\end{align}
where $\ev(t) = \mathrm{exp}\left[-\iu H_{\sm{\mathrm{tot}}}t\right]$ is the total time evolution operator. Assuming that each of the baths are weakly coupled to the system we may expand the total evolution operator up to $2$-nd order to read
\begin{align}
\label{eq:no2.6}
\ev(t)&\approx\e^{-\iu \Hz t}\Biggl\{\id-\sum_{\al}\left[\iu \int_{0}^{t}\drm q \left(\tilde{H}_{{\sm{\mathrm{S}}}\al}(q)+\tilde{H}_{{\sm{\mathrm{R}}}\al}(q)\right) \right.\Biggr.\nonumber \\
&\Biggl.\left. +\int_{0}^{t}\drm q\, \tilde{H}_{{\sm{\mathrm{S}}}\al}(q)\int_{0}^{q} \drm u\, \tilde{H}_{{\sm{\mathrm{S}}}\al}(u)\right]\Biggr\},
\end{align}
where $\Hz = H_{\sm{\mathrm{S}}}+\sum_{\al}H_{\al}$ and all operators with $\sim$'s denote the Heisenberg evolution (also known as free evolution) under $\Hz$, i.e., $\tilde{O}(x)=\e^{\iu \Hz x} O \e^{-\iu \Hz x}$. 

Assuming that the system and the baths are decoupled initially, i.e., $\dm_{\sm{tot}}(0)=\dm_{\sm{\mathrm{S}}}(0)\Pi_{\al}^{\tp}\dm_{\al}(0)$, with each bath being in its canonical distribution, i.e., $\dm_{\al}(0)= \mathrm{exp}\left[-\be^{\al}H_{\al}\right]/Z_{\al}$, and tracing over the bath degrees of freedom we obtain,
\begin{align}
\label{eq:no2.7}
\frac{\drm \dm(t)}{\drm t}&=-\iu\bigl[(H_{\sm{\mathrm{S}}}+\sum_{\al}H_{{\sm{\mathrm{R}}}\al}),\dm(t)\bigr]+\mathcal{R}(t),
\end{align}
where $\dm(t)$ is the RDM of the system and $[\cdot\,,\cdot]$ is the commutator. The \emph{relaxation} operator $\mathcal{R}$, which ensures that the system is damped by the baths, is given by,
\begin{align}
\label{eq:no2.8}
\mathcal{R}(t)&=-\sum_{\al}\int_{0}^{t}\drm q\Bigl\{ \bigl[Y^{\al},\tilde{Y}^{\al}(q-t)\dm(t)\bigr]\ch^{\al}(t-q)\Bigr. \nonumber \\
\Bigl. &-\bigl[Y^{\al},\dm(t)\tilde{Y}^{\al}(q-t)\bigr]\ch^{\al}(q-t)\Bigr\},
\end{align}
where $\ch^{\al}(\tau)=\mathrm{Tr}_{\al} \left[\tilde{B}^{\al}(\tau)B^{\al}\dm_{\al}(0)\right]$ is the $\al$-th bath correlator, where the trace is over the $\al$-th bath. One might argue here that mathematically this average should be with respect to the $0$-th order density matrix of the bath at time $t$. It is here we invoke the physical assumption that the bath is extremely large and cannot be influenced by the system. This leads the bath to remain in its initially prepared state $ \mathrm{exp}\left[-\be^{\al}H_{\al}\right]/Z_{\al}$. The physical assumption, also commonly referred to as the Born approximation, implies that the bath-correlator follows a Kubo-Martin-Schwinger (KMS) condition \cite{Kubo1957, Martin1959}, i.e., $\ch^{\al}(-\tau) = \ch^{\al}(\tau-\iu\be^{\al})$ and the transition rates $\mathrm{Re}[W_{kl}^{\al}]$ follow detailed balance, i.e., $\mathrm{Re}[W_{kl}^{\al}]=\mathrm{exp}[-\be^{\al}\tf_{kl}]\mathrm{Re}[W_{lk}^{\al}]$. In the derivation above we have assumed $\mathrm{Tr}_{\al} \left[B^{\al}\dm_{\al}(0)\right] = 0$, but more generally the term of the form $-\iu\sum_{\al}[Y^{\al},\dm(t)]\mathrm{Tr}_{\al} \left[B^{\al}\dm_{\al}(0)\right]$  must be added to the right hand side of Eq.~(\ref{eq:no2.7}). It is important to note that the effects of multiple baths is additive in the \emph{relaxation} operator $\mathcal{R}(t)$ because our total Hamiltonian did not contain any cross terms having bath-bath correlations. 

Expressing the non-Markovian master equation Eq.~(\ref{eq:no2.7}) in the energy eigenbasis of the system Hamiltonian we find,
\begin{align}
\label{eq:no2.9}
\frac{\drm\dm_{nm}(t)}{\drm t}&=-\iu \tf_{nm}\dm_{nm}(t)+\sum_{kl}\mathcal{R}_{nm}^{kl}\dm_{kl}(t), 
\end{align}
where $\tf_{nm}=E_{n}-E_{m}$ is the difference in energies of the bare system Hamiltonian and the relaxation four tensor $\mathcal{R}_{nm}^{kl}$, which captures the non-Markovian nature, is given by,
\begin{align}
\label{eq:no2.10}
\mathcal{R}_{nm}^{kl}&=\sum_{\al}\Bigl[Y_{nk}^{\al}Y_{lm}^{\al}\left(W_{nk}^{\al}+W_{ml}^{\al *}\right)\Bigr.\nonumber\\
&\Bigl. -\delta_{l,m} \sum_{j} Y_{nj}^{\al}Y_{jk}^{\al}W_{jk}^{\al}-\delta_{n,k} \sum_{j} Y_{lj}^{\al}Y_{jm}^{\al}W_{jl}^{\al *} \Bigr].
\end{align} 
The rates $W_{kl}^{\al}$ take the form
\begin{align}
\label{eq:no2.11}
W_{kl}^{\al} &= \tilde{W}_{kl}^{\al}+\iu\frac{\dkz^{\al}}{2},
\end{align}
where 
\begin{align}
\label{eq:no2.12}
\tilde{W}_{kl}^{\al}&=\int_{0}^{t} \drm\tau\,\e^{-\iu \tf_{kl} \tau} \ch^{\al}(\tau),\\
\label{eq:no2.13}
\dkz^{\al} &=\sum_{k=1}^{\infty}\frac{c_{k,\al}^{2}}{m_{k,\al}\,\w_{k,\al}^{2}}.
\end{align}
The damping kernel at zero time $\dkz^{\al}$ arises from the renormalization part of the Hamiltonian $H_{\mathrm{\sm{R}}\al}$. 
\section{Steady-state nonequilibrium modified Redfield solution}
\label{sec:3}
The quantum master equation (QME) given in Eq.~(\ref{eq:no2.7}) is analogous to the Redfield master equation, but here in the presence of multiple baths. Hence, similar to the Redfield case, the above equation also does not provide the correct steady state solution \cite{Mori2008, Fleming2011, Thingna2012} and contains errors in the diagonal terms of the RDM at the $2$-nd order of system-bath coupling. To avoid these errors and to obtain the RDM correct up to $2$-nd order, in this section, we employ the techniques of the modified Redfield solution \cite{Thingna2012} for the non-trivial nonequilibrium steady-state scenario. 
\subsection{Extracting the correct steady state RDM elements from the quantum master equation}
\label{sec:3.1}
We start by extracting the correct steady-state elements of the RDM from the time-local Redfield-like QME Eq.~(\ref{eq:no2.9}). In order to establish a steady state we first take the limit $t\rightarrow\infty$, by setting the upper limit of the integral in Eq.~(\ref{eq:no2.12}) to $\infty$. The steady state condition then implies
\begin{align}
\label{eq:no3.1}
\frac{\drm\dm_{nm}(t)}{\drm t} &= 0.
\end{align}
Since we assumed a weak system-bath coupling approximation while deriving the time-local Redfield-like QME, we consistently do the same in case of the steady-state RDM $\dm_{nm}$ and assume a general power series expansion in the coupling strength of the form,
\begin{align}
\label{eq:no3.2}
\dm_{nm}&=\sum_{i=0,2,4,\cdots}\dm_{nm}^{\sm{(i)}}.
\end{align}
Above, $\dm^{\sm{(i)}}$ is the $i$-th order RDM, where $i$ indicates the power dependence of the system-bath coupling. The $\dm_{nm}^{\sm{(0)}}$ above should be interpreted as the RDM obtained in the limit the system-bath coupling goes to zero. The limit is crucial to ensure that the system, instead of staying in its initial state, feels the effect of the bath and relaxes to the correct steady state. 

Similar to the equilibrium case \cite{Fleming2011, Thingna2012}, it can be shown that the steady-state diagonal elements obtained by solving Eq.~(\ref{eq:no2.9}) are incorrect in $2$-nd order of system-bath coupling. Hence, using Eq.~(\ref{eq:no3.2}) in Eq.~(\ref{eq:no2.9}) and solving order-by-order we obtain the $0$-th order elements as, 
\begin{align}
\label{eq:no3.3.1}
&\sum_{\al ,k}\left(Y_{nk}^{\al}Y_{kn}^{\al}\Wre_{nk}-\delta_{n,k}\sum_{l}Y_{nl}^{\al}Y_{lk}^{\al}\Wre_{lk}\right)\dm_{kk}^{(0)} = 0,\\
\label{eq:no3.3.2}
&\mathrm{while~for}~(n \neq m),~~~~~~~~~~~~~~~~~~~~~~~~~~~~~\dm_{nm}^{(0)} = 0,
\end{align}
where $\Wre_{kl} = \mathrm{Re}[\tilde{W}_{kl}^{\al}]$. Unlike the equilibrium case, the solution to the above is in general not a Gibbs-distribution due to the lack of a detailed balance condition. The $2$-nd order off-diagonal elements are given by,
\begin{align}
\label{eq:no3.4}
\dm_{nm}^{(2)} & = \frac{1}{\iu\tf_{nm}} \sum_{\al ,k} Y_{nk}^{\al}Y_{km}^{\al} \left[ \Bigl(W_{nk}^{\al}+W_{mk}^{\al*}\Bigr)\dm_{kk}^{(0)} \right. \nonumber\\
& \left.-W_{kn}^{\al*}\dm_{nn}^{(0)}-W_{km}^{\al}\dm_{mm}^{(0)}\right], ~~~~(n \neq m).
\end{align}
The above set of equations describing the $0$-th order and $2$-nd order off-diagonal elements have been obtained under the assumption that our \emph{bare} system Hamiltonian does not posses any degeneracies in its energy eigen-spectrum. The $0$-th order equation, Eq.~(\ref{eq:no3.3.2}), can also be equivalently obtained in the van Hove limit (also sometimes referred to as the Davies limit \cite{Davies1974, Dumcke1979}) by rescaling the time $t$ and coupling strength $\ep$ as $\ep t = \tau$, where $\tau$ should always remain constant \cite{Haake1973}. In the steady state, since $t\rightarrow\infty$, the coupling obeys $\ep\rightarrow 0$ so that $\tau$ remains constant. This causes the $2$-nd order off-diagonal elements to vanish in the steady state and hence we make use of the time-local Redfield-like QME which retains some of the crucial $2$-nd order information.
\subsection{Analytic continuation to obtain $2$-nd order diagonal elements}
\label{sec:3.2}
Following the same reasoning used in our previous work \cite{Thingna2012}, in this section we obtain the $2$-nd order diagonal elements of the steady-state nonequilibrium RDM. In order to do this we make use of analytic continuation techniques and use only the information provided by a $2$-nd order time-local Redfield-like QME Eq.~(\ref{eq:no2.9}). We start by a careful inspection of the $2$-nd order off-diagonal elements Eq.~(\ref{eq:no3.4}) and assume that the energy $E_{m}$ continuously approaches $E_{n}$ by a small complex parameter $z$; i.e., $E_{m}=E_{n}-z$, yielding,
\begin{align}
\label{eq:no3.5}
\dm_{nn}^{(2)} & \propto \lim_{z \to 0}\Biggl\{\frac{1}{\iu z} \sum_{\al ,k} Y_{nk}^{\al}Y_{kn}^{\al} \biggl[ \left(\Wre_{nk}(0)+\Wre_{nk}(-z)\right)\dm_{kk}^{(0)} \biggr.\Biggr. \nonumber\\
\Biggl.\biggl. &-\left(\Wre_{kn}(0)+\Wre_{kn}(z)\right)\dm_{nn}^{(0)}\biggr]\Biggr.\nonumber\\
\Biggl. &+\frac{1}{z} \sum_{\al ,k} Y_{nk}^{\al}Y_{kn}^{\al} \biggl[ \left(\Wim_{nk}(0)-\Wim_{nk}(-z)\right)\dm_{kk}^{(0)} \biggr.\Biggr. \nonumber\\
\Biggl.\biggl. &-\left(\Wim_{kn}(0)-\Wim_{kn}(-z)\right)\dm_{nn}^{(0)}\biggr.\Biggr. \nonumber \\
\Biggl.\biggl. &+\left(\Wim_{kn}(-z)+\frac{\dk_{\sm{0}}}{2}\right)z\frac{\partial\dm_{nn}^{(0)}}{\partial E_{n}}\biggr]\Biggr\},
\end{align}
where,
\begin{align}
\label{eq:no3.6}
\tilde{W}_{kl}^{\al}(z)&=\int_{0}^{\infty}\drm\tau\,\e^{-\iu(\tf_{kl}+z)\tau}\ch^{\al}(\tau),
\end{align}
$\tilde{W}_{kl}^{\al\p}(z) = \mathrm{Re}[\tilde{W}_{kl}^{\al}(z)]$, and $\tilde{W}_{kl}^{\al\p\p}(z) = \mathrm{Im}[\tilde{W}_{kl}^{\al}(z)]$. Above, since $\dm_{mm}^{(0)}$ also depends on the energy $E_{m}$, we made use of the Taylor expansion of $\dm_{mm}^{(0)}$ of the form,
\begin{align}
\label{eq:no3.7}
\lim_{E_{m} \to E_{n}}\dm_{mm}^{(0)}&\simeq\dm_{nn}^{(0)}+z\frac{\partial\dm_{nn}^{(0)}}{\partial E_{n}}.
\end{align}
Noting that $\lim_{z\to 0}\tilde{W}_{kl}^{\al}(z)=\tilde{W}_{kl}^{\al}(-z)=\tilde{W}_{kl}^{\al}$ we find,
\begin{align}
\label{eq:no3.8}
\dm_{nn}^{(2)} & \propto \sum_{\al ,k} Y_{nk}^{\al}Y_{kn}^{\al} \left[ V_{nk}^{\al\p\p}\dm_{kk}^{(0)}-V_{kn}^{\al\p\p}\dm_{nn}^{(0)}\right]+W_{kn}^{\al\p\p}\frac{\partial\dm_{nn}^{(0)}}{\partial E_{n}},\nonumber\\
\end{align}
where
\begin{align}
\label{eq:no3.9}
V_{kl}^{\al\p\p} & = \frac{\partial\Wim_{kl}}{\partial\tf_{kl}} = \lim_{z\to 0}\frac{\Wim_{kl}(0)-\Wim_{kl}(-z)}{z},
\end{align}
$W_{kl}^{\al\p\p} = \mathrm{Im}[W_{kl}^{\al}]$, and $\Wim_{kl} = \mathrm{Im}[\tilde{W}_{kl}^{\al}]$. In order to obtain Eq.~(\ref{eq:no3.8}) we have omitted one of the terms which takes the same form as the left hand side of Eq.~(\ref{eq:no3.3.1}), but is zero in the limit $z\rightarrow 0$. The limiting procedure above is independent of the way in which $E_{m}$ approaches $E_{n}$ implying the uniqueness of the limit and thus physically the steady state.

A reader might have observed the proportionality signs in Eqs.~(\ref{eq:no3.5}) and (\ref{eq:no3.8}). This is mainly because the diagonal elements of the RDM have an additional constraint of normalization. At the $0$-th order the trace should be unity, which immediately implies that the $2$-nd order elements should be traceless. Equation~(\ref{eq:no3.8}) does not preserve this normalization condition, due to the analytic continuity procedure. Therefore, we re-normalize the RDM using,
\begin{align}
\label{eq:no3.10}
\dm_{nn}&=\frac{\dm_{nn}^{(0)}+\dm_{nn}^{(2)}}{\sum_{k}(\dm_{kk}^{(0)}+\dm_{kk}^{(2)})} \nonumber \\
&\simeq\dm_{nn}^{(0)}+\dm_{nn}^{(2)}-\dm_{nn}^{(0)}\sum_{k}\dm_{kk}^{(2)},
\end{align}
where we have ignored the $4$-th and higher order terms and used the condition $\sum_{k}\dm_{kk}^{(0)} = 1$. Therefore, after this normalization, Eq.~(\ref{eq:no3.8}) transforms into,
\begin{align}
\label{eq:no3.11}
\dm_{nn}^{(2)} & = \sum_{\al ,k} Y_{nk}^{\al}Y_{kn}^{\al} \left[V_{nk}^{\al\p\p}\dm_{kk}^{(0)}-V_{kn}^{\al\p\p}\dm_{nn}^{(0)}+W_{kn}^{\al\p\p}\frac{\partial\dm_{nn}^{(0)}}{\partial E_{n}}\right] \nonumber \\
&-\dm_{nn}^{(0)}\sum_{\al ,k,l}Y_{lk}^{\al}Y_{kl}^{\al}W_{kl}^{\al\p\p}\frac{\partial\dm_{ll}^{(0)}}{\partial E_{l}},
\end{align}
where $\partial\dm_{nn}^{(0)}/\partial E_{n}$ can be obtained by differentiating Eq.~(\ref{eq:no3.3.1}) as,
\begin{align}
\label{eq:no3.12}
\frac{\partial\dm_{nn}^{(0)}}{\partial E_{n}} & = \frac{\sum_{\al ,\substack{k\ne n}}Y_{nk}^{\al}Y_{kn}^{\al}\left(V_{nk}^{\al\p}\dm_{kk}^{(0)}+V_{kn}^{\al\p}\dm_{nn}^{(0)}\right)}{\sum_{\al ,\substack{k\ne n}}Y_{nk}^{\al}Y_{kn}^{\al}\Wre_{kn}},
\end{align}
where $V_{kl}^{\al\p} = \partial\Wre_{kl}/\partial\tf_{kl}$. The above $2$-nd order diagonal elements constitute the main result of this paper. If we compare the above result to the equilibrium case \cite{Thingna2012}, we find that the only difference amounts to an extra summation index $\al$. This is obvious in hindsight because our initial model did not have any bath-bath correlations. Thus, Eqs.~(\ref{eq:no3.3.1}),~(\ref{eq:no3.3.2}),~(\ref{eq:no3.4}), and~(\ref{eq:no3.11}) form our nonequilibrium modified Redfield solution, which represents the RDM correct up to $2$-nd order in the system-bath coupling strength. 

Although for a general system one needs to solve Eq.~(\ref{eq:no3.3.1}) using numerical techniques, we would like to point out one special regime where we may be able to obtain the solution analytically. Let us consider the regime where the temperature differences are small and identical baths with different temperatures are connected to a single system degree of freedom, i.e., $Y^{\al} = Y $ for all $\al$. In this special regime an approximate Gibbs-distribution like state exists for the $0$-th order RDM, i.e., $\dm^{(0)} \approx \mathrm{exp}(-\bar{\be}H_{\mathrm{\sm{S}}})/\mathrm{Tr_{\sm{S}}}\left[\mathrm{exp}(-\bar{\be}H_{\mathrm{\sm{S}}})\right]$, where $\bar{\be}$ is the inverse of the arithmetic average temperature of various baths. Such a solution allows us to obtain the $2$-nd order RDM analytically, since the $2$-nd order terms are expressions which depend on $\dm^{(0)}$, refer Eqs.~(\ref{eq:no3.4}) and~(\ref{eq:no3.11}). The existence of approximate Gibbs-distributions have been investigated before for various classical models \cite{Lebowitz1959} and are valid in the quantum regime as long as the energy spectrum has a finite width. The width of the spectrum plays an important role and the narrower the energy spectrum the wider the regime of validity of the approximate Gibbs state (for more information see Appendix). Although such a simple manifestation is true for the $0$-th order RDM the statement can not be extended to higher orders in terms of the generalized Gibbs distribution, i.e., $\dm \neq \mathrm{Tr}_{\mathrm{\sm{B}}}[\mathrm{exp}(-\bar{\be}H_{\mathrm{\sm{tot}}})]/\mathrm{Tr}[\mathrm{exp}(-\bar{\be}H_{\mathrm{\sm{tot}}})]$.
\section{Corroboration and comparison for the quantum harmonic oscillator}
\label{sec:4}

In this section we compare our nonequilibrium modified Redfield solution to the lone exact result of the quantum harmonic oscillator obtained via techniques of nonequilibrium Green's function \cite{Dhar2012}. For this specific case we will choose our system Hamiltonian to take the form,
\begin{align}
\label{eq:no4.1}
H_{\mathrm{\sm{S}}} & = \frac{p^{2}}{2M} + \frac{1}{2}M\w_{\sm{0}}^{2}x^{2},
\end{align} 
where $x$, $p$, $M$, and $\w_{\sm{0}}$ are the position, momentum, mass and angular frequency of the oscillator, respectively. The harmonic oscillator poses a tough numerical challenge for traditional QMEs like Eq.~(\ref{eq:no2.9}) because of the relaxation four tensor $\mathcal{R}_{nm}^{kl}$ which scales as $N^4$, where $N$ is the system Hilbert space dimension. The memory requirement for these traditional QMEs with a modest $N = 40$ is approximately 40 MB for storing only the relaxation tensor $\mathcal{R}_{nm}^{kl}$. Since our technique does not need to store the relaxation tensor and only relies on the storage of the rates $W_{kl}^{\al}$, which scale as $N^{2}$, the memory requirement drastically drops to a mere 25 kB. This enables us to deal with large system Hilbert spaces like that of the harmonic oscillator. Also \emph{uncontrolled} approximations like the rotating wave approximation \cite{Blum1996} should not be applied to this model because of the equi-spaced energy spectrum, making this example a viable testing ground.

We then couple the harmonic oscillator linearly to the minimal transport setup involving two baths ($\al = \mathrm{L,R}$) via the position coupling, i.e, $Y^{\sm{\mathrm{L,R}}}=x$ and $B^{\al}=-\sum_{k=1}^{\infty}c_{k,\al}x_{k,\al}$ in Eq.~(\ref{eq:no2.4}). In order to describe the baths we will make use of the spectral density $\sd^{\al}(\w)$ defined as,
\begin{align}
\label{eq:no4.2}
\sd^{\al}(\w)&=\pi\sum_{k=1}^{\infty}\frac{c_{k,\al}^{2}}{2m_{k,\al}\,\w_{k,\al}}\delta(\w-\w_{k,\al}).
\end{align}
Now we will choose both baths to have same parameters, i.e., $\sd^{\mathrm{\sm{L}}}(\w)=\sd^{\mathrm{\sm{R}}}(\w)=\sd(\w)$, which we will choose to be of the form,
\begin{align}
\label{eq:no4.3}
\sd(\w) & = \frac{M\dk \w}{1+\left(\w/\w_{\sm{\mathrm{D}}}\right)^{2}}.
\end{align}
The above form of the spectral density is known as the Lorentz-Drude form, where $\w_{\sm{\mathrm{D}}}$ denotes the cut-off frequency and $\dk\propto \sum_{k=1}^{\infty} c_{k}^{2}$ is the phenomenological Stokesian damping coefficient which characterizes the system-bath coupling strength. 

Using this definition of spectral density we can now recast the bath correlator $\ch^{\al}(\tau)$ as,
\begin{align}
\label{eq:no4.4}
\ch^{\al}(\tau)=\int_{0}^{\infty}\frac{\drm\w}{\pi}\,\sd(\w) \Biggl[&\mathrm{coth}\left(\frac{\be^{\al}\w}{2}\right)\mathrm{cos}(\w\tau)\Biggr.\nonumber\\
\Biggl. &-\iu \mathrm{sin}(\w\tau)\Biggr].
\end{align}
For the given Lorentz-Drude model the bath correlator can be evaluated analytically and it takes the form,
\begin{align}
\label{eq:no4.5}
\ch^{\al}(\tau)&=\frac{M\dk}{2}\w_{\sm{\mathrm{D}}}^{2}\e^{-\w_{\sm{\mathrm{D}}}\tau}\left[\mathrm{cot}\left(\frac{\be^{\al} \w_{\sm{\mathrm{D}}}}{2}\right)-\iu\rm{sgn}(\tau)\right]\nonumber\\
&-\frac{2M\dk}{\be^{\al}}\sum_{j=1}^{\infty}
\frac{\nu_{j}^{\al}\e^{-\nu_{j}^{\al}\tau}}{1-(\nu_{j}^{\al}/\w_{\sm{\mathrm{D}}})^{2}},
\end{align}
where $\nu_{j}^{\al} = 2\pi j/\be^{\al}$ are known at the Matsubara frequencies. The damping kernel defined in Eq.~(\ref{eq:no2.13}) can also be written in terms of $\sd(\w)$ as,
\begin{align}
\label{eq:no4.6}
\dkz^{\mathrm{\sm{L}}}&=\dkz^{\mathrm{\sm{R}}}=\dkz =\frac{2}{\pi}\int_{0}^{\infty}\drm\w \,\frac{\sd(\w)}{\w},
\end{align}
and for the Lorentz-Drude model $\dkz=\dk\w_{\sm{\mathrm{D}}}$.

\begin{figure}
\begin{center}
\includegraphics[scale=0.33]{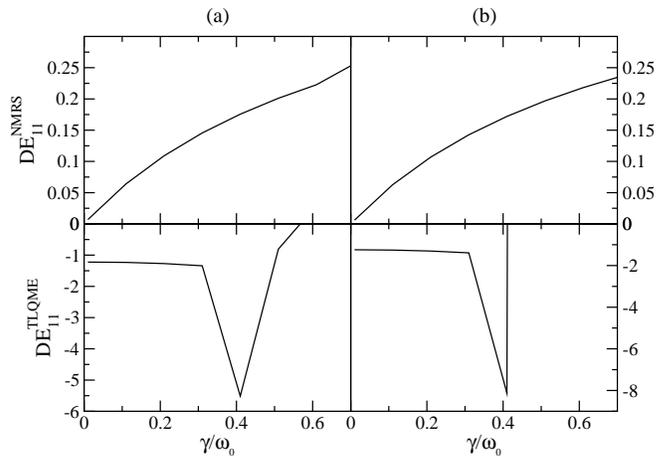}
\end{center}
\caption{Plot of the discrepancy error $\mathrm{DE}^{\sm{\mathrm{X}}}_{11}$, see Eq.~(\ref{eq:no4.9}), of the ground state population versus the dimensionless system-bath coupling strength ($\dk/\w_{\sm{0}}$) for a quantum harmonic oscillator connected to two heat baths. Top panel shows the discrepancy error for the nonequilibrium modified Redfield solution ($X =$ NMRS) and the bottom panel is for the time-local Redfield-like quantum master equation ($X =$ TLQME). Figure (a) is for temperatures $T_{\mathrm{\sm{L}}} = 156$K and $T_{\mathrm{\sm{R}}} = 140$K, whereas Figure (b) is for $T_{\mathrm{\sm{L}}} = 156$K and $T_{\mathrm{\sm{R}}} = 78$K. Other parameters used for the calculation are: $M$ = 1u, $\w_{\sm{0}} = 1.3\times10^{14}$Hz, and $\w_{\mathrm{\sm{D}}} = 10\w_{\sm{0}}$.}
\label{fig:1}
\end{figure}

Therefore, the components of the rates $\tilde{W}^{\al}$, defined in Eq.~(\ref{eq:no2.12}), read
\begin{align}
\label{eq:no4.7}
\Wre_{kl} &=\sd(\tf_{kl})\,n^{\al}(\tf_{kl}),\\
\label{eq:no4.8}
\Wim_{kl} &=\frac{M\dk \w_{\sm{\mathrm{D}}}^{2}\tf_{lk}}{2(\w_{\sm{\mathrm{D}}}^{2}+\tf_{kl}^{2})}\left[\mathrm{cot}\left(\frac{\be^{\al} \w_{\sm{\mathrm{D}}}}{2}\right)+\frac{\w_{\sm{\mathrm{D}}}}{\tf_{kl}}\right] \nonumber \\
&+\frac{2 M\dk}{ \tf_{kl}\be^{\al}}\sum_{j=1}^{\infty}\frac{\nu_{j}^{\al}}{(1-(\nu_{j}^{\al}/\w_{\sm{\mathrm{D}}})^{2})(1+(\nu_{j}^{\al}/\tf_{kl})^{2})},
\end{align}
where the Bose--Einstein distribution function $n^{\al}(\tf_{kl}) = \left[\mathrm{exp}(\be^{\al}\tf_{kl})-1\right]^{-1}$, with the inverse temperature $\be^{\al}$ for each bath.

Now once we have defined the bath properties and the coupling to the system we calculate the nonequilibrium modified Redfield solution for the harmonic oscillator problem using a fixed number of energy levels. We truncate the number of levels by ensuring that the highest few are unoccupied up to temperatures of $5\times T_{\sm{\mathrm{D}}}$, with $T_{\sm{\mathrm{D}}} = (\h\w_{\sm{0}})/k_{\sm{\mathrm{B}}}$ being the Debye temperature. This results in the use of $\approx 40$ energy levels for the single quantum harmonic oscillator. In order to corroborate with the exact nonequilibrium Green's function (NEGF) results of Dhar \emph{et~al.} \cite{Dhar2012}, we define a discrepancy error:
\begin{align}
\label{eq:no4.9}
\mathrm{DE}^{\sm{\mathrm{X}}}_{kl} &\equiv \left[\dm^{\sm{\mathrm{NEGF}}}_{kl}-\dm^{\sm{\mathrm{X}}}_{kl}\right]/(\dk/\w_{\sm{0}}),
\end{align}
where $\dm^{\sm{\mathrm{NEGF}}}$ describes the exact RDM obtained via the NEGF method and $\dm^{\sm{\mathrm{X}}}$ could be the RDM either from the nonequilibrium modified Redfield solution ($\mathrm{X} =$ NMRS) or the time-local Redfield-like quantum master equation ($\mathrm{X} =$ TLQME) described in Sec.~\ref{sec:2}. Now, because the $2$-nd order RDM, i.e., $\dm^{(2)}$, is proportional to $\dk$ it is clear that if $\dm^{\sm{\mathrm{NEGF}}}$ matches $\dm^{\sm{\mathrm{X}}}$ up to $2$-nd order then the discrepancy error $\mathrm{DE}^{\sm{\mathrm{X}}} \rightarrow 0$ as $\dk \rightarrow 0$. In other words, $\dm^{\sm{\mathrm{NEGF}}}$ matches $\dm^{\sm{\mathrm{X}}}$ in first order of dissipation strength $\gamma$ ($2$-nd order of coupling strength) for arbitrary value of dissipation if only $\mathrm{DE}^{\sm{\mathrm{X}}} \rightarrow 0$ as $\dk \rightarrow 0$ is obeyed.

In Fig.~\ref{fig:1} we depict the discrepancy error $\mathrm{DE}^{\sm{\mathrm{X}}}_{11} $ in the first level population of the RDM. Since the temperatures of the baths are kept low the first level populations depict a fair representation of the entire RDM. Clearly the discrepancy error shows the correct behaviour only for the nonequilibrium modified Redfield solution (NMRS), Fig.~\ref{fig:1} top, whereas for the time-local Redfield-like QME (TLQME), Fig.~\ref{fig:1} bottom, as $\dk \rightarrow 0$ the discrepancy error goes to a constant. This indicates that the TLQME contains errors in the $2$-nd order of the RDM. These errors in nonequilibrium can lead to inaccurate results, especially when one tries to calculate the current based on the local operator definition. Thus it is only the NMRS which is well suited for such applications since it accurately captures all system-bath coupling effects to the lowest order. Importantly, the bath temperatures do not play a major role in Fig.~\ref{fig:1} and same qualitative behaviour is observed for all temperature ranges and differences.  
\begin{figure}
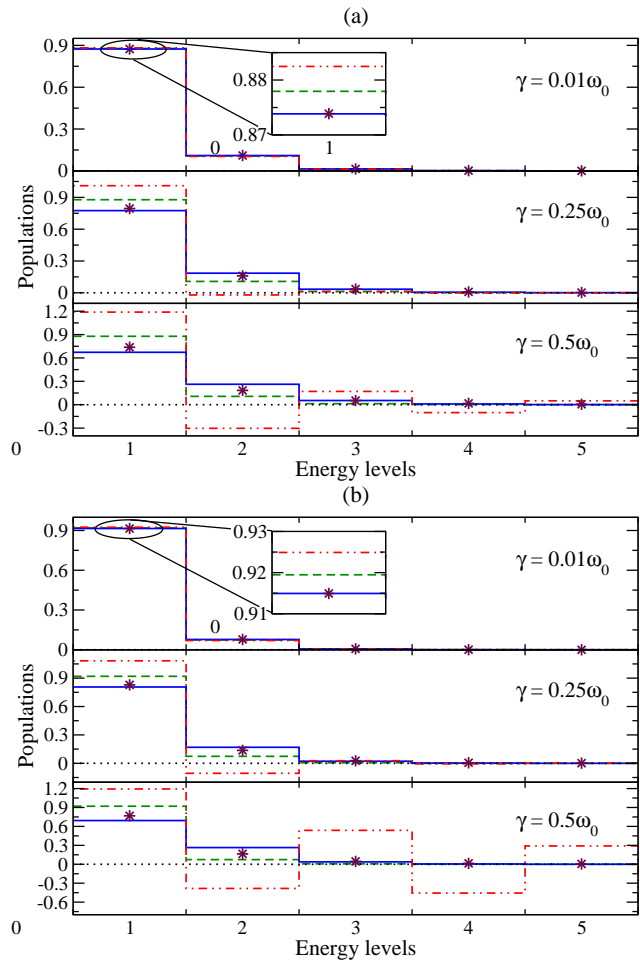

\begin{center}
\includegraphics[scale=0.35]{fig2a.eps}
\includegraphics[scale=0.35]{fig2b.eps}
\end{center}
\caption{(Color online) Histogram of the populations for the first five lowest lying energy levels for different system-bath coupling strengths for a damped quantum harmonic oscillator. Figure~(a) corresponds to $T_{\mathrm{\sm{L}}} = 312$K, $T_{\mathrm{\sm{R}}} = 280.8$K and in Fig.~(b) $T_{\mathrm{\sm{L}}} = 312$K, $T_{\mathrm{\sm{R}}} = 156$K. Inset in top panels is a zoom in of the first energy level populations. The solid (blue) lines correspond to our nonequilibrium modified Redfield solution (NMRS), the dash-dotted (red) lines present the results for the time-local Redfield-like quantum master equation (TLQME), the dashed (green) lines depict the results for the Lindblad-like solution and the (maroon) stars represent the exact NEGF results. The parameters used for the calculation are $M = 1~\mathrm{u}$ , $\w_{\sm{0}} = 1.3\times10^{14}~\mathrm{Hz}$, and $\w_{\sm{\mathrm{D}}} = 10 ~\w_{\sm{0}}$.}
\label{fig:2}
\end{figure}

Next, in Fig.~\ref{fig:2} we compare our nonequilibrium modified Redfield solution (NMRS) to the time-local Redfield-like QME (TLQME), the Lindblad-like master equation \cite{Lindblad1976, Alicki1987, Carmichael1994} and the exact NEGF results \cite{Dhar2012}. The Lindblad-like solution (dashed green line) is (completely) positive, which has been critiqued before for a system connected to a single bath \cite{Pechukas1994, Fonseca2004, Shaji2005}, and it fails to capture the effects of the finite system-bath coupling. These erroneous behaviours of the Lindblad-like solution could present a serious drawback to tackle transport problems where the dependence on coupling strength is of primal importance. On the other hand, the TLQME (dash-dotted red line) produces unphysical negative probabilities even for moderate coupling strengths, Fig.~\ref{fig:2}: middle panel $\dk/w_{\sm{0}} = 0.25$ and bottom panel $\dk/w_{\sm{0}} = 0.5$. In the equilibrium case this problem has been critiqued repeatedly \cite{Munro1996, Blanga1996}, but to the best of our knowledge the issue has not been addressed in the nonequilibrium scenario. Clearly the breaking of positivity for the TLQME is a result of incorrect $2$-nd order diagonal elements, because our NMRS (solid blue line) seems to behave reasonably well for coupling strengths well beyond the naive expectation for a perturbative master equation. Also, as compared to the exact NEGF results (maroon stars), which take into account all orders of coupling strength, our NMRS result shows excellent agreement even in the moderate coupling strength regime, i.e., for $\dk/w_{\sm{0}} = 0.25$ and $\dk/w_{\sm{0}} = 0.5$. In this moderate coupling strength regime it is expected that higher orders of the coupling strength will also play a role, due to which a small difference is observed between the exact NEGF results and our second order NMRS approach. It should also be noted that the NMRS is not (completely) positive and can even give rise to negative populations if the coupling strength increases far beyond its \emph{a priori} regime of validity of finite weak coupling. The qualitative features of the results described above do not depend on temperature differences (as seen from Figs.~\ref{fig:2}a and~\ref{fig:2}b) or absolute temperature, implying that our NMRS is an excellent method to accurately capture nonequilibrium effects in the weak to moderate system-bath coupling regimes.
\section{Concluding remarks and future directions}
\label{sec:5}
In summary, we presented a novel technique based on analytic continuity to evaluate the steady-state reduced density matrix of a general anharmonic system connected to multiple heat baths correct up to $2$-nd order in the system-bath coupling. Our novel nonequilibrium modified Redfield solution (NMRS) was verified against the only known exact nonequilibrium solution of the quantum harmonic oscillator and excellent agreement is obtained between these two approaches. Other ``popular'' quantum master equations were then compared against our NMRS and considerable differences were found in the regime of moderate system-bath coupling. In this regime, it was only the NMRS that provides physically reliable solutions whereas the other approaches either violated positivity or did not change with increasing coupling strength. In order to study systems in nonequilibrium the moderate (or at least weak but finite) coupling strength regime is extremely crucial because some of the most interesting phenomena, like transport, solely depend on the strength of the coupling and are trivially zero for vanishing couplings. Thus, in cases where local current operators are defined our NMRS presents an accurate non-phenomenological approach to deal with steady-state transport.

Even though our approach is accurate and numerically efficient, several unresolved challenges still remain. One subtle issue lies in dealing with systems which posses a degeneracy for the eigenvalues in the bare system-Hamiltonian. One can mathematically circumvent this issue by re-calculating the order-by-order solution for degenerate systems, as done in Sec.~\ref{sec:3.1} for the case of non-degenerate systems, and then use our analytic continuity approach to tackle the $2$-nd order diagonal elements correctly. Another important challenge lies in the hierarchical nature of the master equations, i.e., in order to know the $n$-th order RDM one requires a $n+2$-th order master equation. Our novel approach has demonstrated that up to $2$-nd order there is a reasonable route to bypass this hierarchical problem and work at a given order, but it is still an open question if such a scheme would even work for higher orders. A deeper mathematical or physical understanding of why the analytic continuation works is also not settled. It is also not clear how one could extend our scheme to study the relaxation dynamics. The uniqueness of the steady state makes our approach feasible, but the dynamical problem is an herculean task because there could be various equivalent dynamical routes. Despite its limitation to steady state we are confident that our approach paves a new way to address nonequilibrium physics in general anharmonic systems beyond the vanishing coupling limit.
\section*{Acknowledgements}
\label{sec:6}
The authors like to thank Peter Talkner and Hangbo Zhou for fruitful discussions.
\appendix*
\section{Approximate Gibbs-distribution in the limit of vanishing coupling}
\label{append:A}
\begin{figure}
\begin{center}
\includegraphics[scale=0.52]{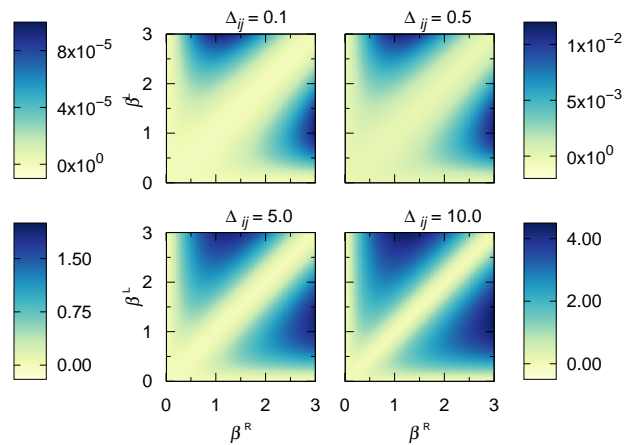}
\end{center}
\caption{(Color Online) Plot of the $\chi\left(\tf_{ij},\be^{\mathrm{\sm{L}}},\be^{\mathrm{\sm{R}}}\right)$ as a function of inverse temperatures of the baths $\be^{\mathrm{\sm{L}}}$ and $\be^{\mathrm{\sm{R}}}$ for different energy differences $\tf_{ij}$.}
\label{fig:3}
\end{figure}
In specific parameter regimes it is possible to approximate the $0$-th order reduced density matrix (RDM), described in the main body of the paper, as an effective canonical distribution $\dm^{(0)} \approx \e^{-\bar{\be}H_{\mathrm{\sm{S}}}}/\mathrm{Tr_{\sm{S}}}\left[\e^{-\bar{\be}H_{\mathrm{\sm{S}}}}\right]$, with an effective inverse temperature. Here, we would like to numerically illustrate this idea using a simple example of a system connected to two identical harmonic heat baths, with different temperatures, denoted by ``L'' (left bath) and ``R'' (right bath). We look at the equation describing the $0$-th order RDM, Eq.~(\ref{eq:no3.3.1}), and limit our investigation to the regime where both the baths are connected to the same system operator, i.e., $Y^{\mathrm{\sm{L}}} = Y^{\mathrm{\sm{R}}} = Y$. Therefore Eq.~(\ref{eq:no3.3.1}) can be recast as,
\begin{align}
\label{eq:noA.1}
\sum_{k}\left(Y_{nk}Y_{kn}\tilde{W}_{nk}^{c\p}-\delta_{n,k}\sum_{l}Y_{nl}Y_{lk}\tilde{W}_{lk}^{c\p}\right)\dm_{kk}^{(0)} &= 0,
\end{align}
where $\tilde{W}^{c\p} = \tilde{W}^{\mathrm{\sm{L}}\p}+\tilde{W}^{\mathrm{\sm{R}}\p}$. Eq.~(\ref{eq:noA.1}) resembles an approximate detailed balance equation if the rates $\tilde{W}^{c\p}$ follow 
\begin{align}
\label{eq:noA.2}
\tilde{W}_{ij}^{c\p}&\approx\mathrm{exp}(-\bar{\be}\tf_{ij})\tilde{W}_{ji}^{c\p},
\end{align}
where $\tf_{ij} = E_i - E_j$ is the energy difference of the system Hamiltonian and $\bar{\be} = 2\be^{\mathrm{\sm{L}}}\be^{\mathrm{\sm{R}}}/\left(\be^{\mathrm{\sm{L}}}+\be^{\mathrm{\sm{R}}}\right)$ represents the inverse of the average temperature. Now since both the baths have the same physical properties, i.e., $\sd^{\mathrm{\sm{L}}}(\w)=\sd^{\mathrm{\sm{R}}}(\w)=\sd(\w)$, then using Eq.~(\ref{eq:no4.7}) from the main text, which in fact is true for all spectral densities, it can be shown that Eq.~(\ref{eq:noA.2}) is equivalent to
\begin{align}
\label{eq:noA.3}
&\chi\left(\tf_{ij},\be^{\mathrm{\sm{L}}},\be^{\mathrm{\sm{R}}}\right) =\nonumber\\ &\mathrm{ln}\left[\frac{\e^{\be^{\mathrm{\sm{L}}}\tf_{ij}}+\e^{\be^{\mathrm{\sm{R}}}\tf_{ij}}-2}{2\e^{\left(\be^{\mathrm{\sm{L}}}+\be^{\mathrm{\sm{L}}}\right)\tf_{ij}}-\e^{\be^{\mathrm{\sm{L}}}\tf_{ij}}-\e^{\be^{\mathrm{\sm{R}}}\tf_{ij}}}\right]+\bar{\be}\tf_{ij} = 0
\end{align}

Thus, without the need of defining a system Hamiltonian or the bath properties it is possible to numerically check the validity of Eq.~(\ref{eq:noA.3}) for various energy differences and temperatures as shown in Fig.~\ref{fig:3}. Clearly when $\be^{\mathrm{\sm{L}}} \approx \be^{\mathrm{\sm{R}}}$, i.e., slightly off the diagonals of Fig.~\ref{fig:3}, Eq.~(\ref{eq:noA.3}) is satisfied to a large extent. In cases where the energy difference of the system Hamiltonian is not that large the regime of validity of the approximate detailed balance condition goes well beyond the small temperature-difference regime. Keeping in mind that the Eq.~(\ref{eq:noA.3}) should be valid for all combinations of energy differences we infer that the approximate Gibbs behaviour, beyond the trivial small temperature-difference regime, is applicable for systems whose width of the energy spectrum is much smaller than the temperatures of the baths. 

\end{document}